\documentclass[aps,twocolumn,superscriptaddress]{revtex4}
\usepackage{graphicx}

\begin{document}
\title{Finite size effects on kaonic pasta structures}

\author{Toshiki Maruyama}
\email{maruyama.toshiki@jaea.go.jp}
\affiliation{Advanced Science Research Center, Japan Atomic Energy Research Institute\footnote{Present name: Japan Atomic Energy Agency}, Tokai, Ibaraki 319-1195, Japan}
\author{Toshitaka Tatsumi}
\email{tatsumi@ruby.scphys.kyoto-u.ac.jp}
\affiliation{Department of Physics, Kyoto University, Kyoto, 606-8502, Japan}
\author{Dmitri~N.~Voskresensky}
\email{D.Voskresensky@gsi.de}
\affiliation{Moscow Institute for Physics and Engineering, Kashirskoe sh.~31, Moscow 115409, Russia}
\affiliation{Gesellschaft
 f\"ur Schwerionenforschung mbH, Planckstr. 1,
64291 Darmstadt, Germany}
\author{Tomonori Tanigawa}
\affiliation{Japan Society for the Promotion of Science, Tokyo 102-8471, Japan}
\affiliation{Advanced Science Research Center, Japan Atomic Energy Research Institute\footnote{Present name: Japan Atomic Energy Agency}, Tokai, Ibaraki 319-1195, Japan}
\author{Tomoki Endo}
\email{endo@ruby.scphys.kyoto-u.ac.jp}
\affiliation{Department of Physics, Kyoto University, Kyoto, 606-8502, Japan}
\author{Satoshi Chiba}
\email{chiba.satoshi@jaea.go.jp}
\affiliation{Advanced Science Research Center, Japan Atomic Energy Research Institute\footnote{Present name: Japan Atomic Energy Agency}, Tokai, Ibaraki 319-1195, Japan}

\date{\today}

\begin{abstract}
Non-uniform structures of mixed phases at the first-order phase
transition to charged kaon condensation are
studied using a density functional theory within the
relativistic mean field model. 
Including electric field effects and applying the Gibbs conditions in a proper way, 
we numerically determine density profiles of nucleons, electrons and condensed kaons. 
Importance of charge screening effects is
elucidated and thereby we show that the Maxwell construction is
effectively justified. 
Surface effect is also studied to figure out 
its effect on the density profiles.
\end{abstract}

\maketitle

\section{Introduction}
Concerning phase transitions in nuclear matter, various scenarios have
been discussed, like e.g.\ liquid-gas transition \cite{Gupta}, 
pion condensation \cite{MSTV90},
hadron-quark deconfinement transition
\cite{gle92,HPS93,pei95,voskre,voskre1,dubna,emaru1},  transitions between
different color superconducting phases \cite{ARRW,bed,RR04,BCR03}, quark ferromagnetic 
transition \cite{T05}, charged $\rho$ meson condensation \cite{V97,KV03}, etc., 
and kaon condensation. 
In most cases these are first-order phase transitions (FOPTs).

References \cite{KN86,BLRT20,MTT93,T95,PREPL00,RBP00,GS99,CG00,CGS00,NR00,YT00}
and many others studied possibilities of $s$-wave kaon
condensations, whereas Refs.\ \cite{KVK95,M01}  considered also
those of $p$-wave condensations. 
It has been argued (see \cite{KN86,BLRT20,MTT93,T95,KVK95,PREPL00,RBP00,GS99,CG00,CGS00,NR00,YT00,M01} 
and references therein) that the FOPT to a $K^-$ condensate state  might occur in neutron
stars already at densities only several times larger than 
the nuclear saturation density $\rho_0$. 

The possibility of the FOPT to the kaon condensate  state may  lead to
interesting consequences for  physics of neutron stars, such as a delayed
collapse of a supernova to a low-mass black hole,  
a fast cooling mechanism of neutron stars due to the opening up of the
nucleon direct Urca process under a background of the charged kaon condensate, 
etc \cite{T95,Lee96,Prakash97,bb94,bkpp88,t88,fmtt94,pb90}.

Reference \cite{gle92} argued that in  systems with different charged species 
the structured mixed phases might appear due to  the FOPT. 
For example in case of the  nuclear matter below the saturation density, where
the liquid-gas FOPT is relevant, the ``nuclear pasta'' structures
have been studied by many authors
\cite{dubna,Rav83,Has84,Wil85,Oya93,Lor93,Cheng97,Mar98,kido00,Gen00,Gen02,Gen03,maru1,maru05}.

The physical reason for the possibility of the mixed (pasta) phases with
different geometrical structures is that the charge neutrality may hold  only globally
rather than locally as in the Maxwell construction.
Mechanically, a balance between the Coulomb force and the surface tension is 
responsible for  the occurrence of the spatially non-uniform structures.

The appearance of the pasta structures could have important consequences
for the various neutron star phenomena. 
If appeared at an early stage of the neutron star evolution the pasta could 
cause a drastic change of the neutrino opacity. 
It may influence the subsequent  neutron star cooling,
it affects the matter resistance to the stress and consequently the glitch phenomena, etc.

Long time there  existed a naive view that not all the Gibbs conditions can
be satisfied in a description by the Maxwell construction
\cite{GS99,CG00,CGS00,gle92}, 
because the local charge neutrality is implicitly assumed in it.
As the result of previous works, cf.\ \cite{GS99,CG00,CGS00},
it was suggested  that  a broad region of kaonic pasta as a mixed phase 
may occur in neutron stars with various geometrical structures
of the kaon condensed phase (dense phase) embedded in 
usual nuclear matter phase (dilute phase).
However, in recent papers \cite{voskre,voskre1,dubna,emaru1,mtctv,maruKaon}
we have demonstrated that both treatments using the Gibbs conditions and
the Maxwell construction are actually in  agreement with each
other, if one properly includes electric field effects. 
The bare Coulomb field is screened by rearranged charged particles.
Taking first the hadron-quark phase transition as an explicit example
\cite{voskre,voskre1,emaru1}, we showed that with screening
effects included the most realistic parameter choice (density,
chemical potentials, surface tension, etc.) does not allow the
mixed phase due to the mechanical instability of structures. 
Thus if exists, the density region of
the structured mixed phase should be largely limited.

In subsequent papers \cite{maru1,maru05,dubna}, we have explored the 
charge screening effect on the nuclear pasta phases at sub-nuclear densities.
We compared the full calculation with  those either 
not including the screening at all or including
it only non-systematically, as has been performed in previous works.
We found that the screening effect, being systematically treated, 
significantly affects the density range of the nuclear pasta with various geometries,
while it results in only minor corrections  to the EOS. 
We have also emphasized that the proton spatial rearrangement effect resulting in a
screening is much more efficient, compared with the electron screening, because the proton
Debye screening length is essentially smaller than the electron one. 
However, since the charge density as well as the baryon number
density are rather low in the nuclear pasta case, the screening effect 
is still not so prominent compared to that we find for denser systems.

In this paper we continue our research
\cite{voskre,voskre1,emaru1,maru1,maru05,dubna} of the special
role of the charge screening effects in the problem of the
construction of the mixed phases, taking $s$-wave kaon condensation as a
further example, cf.\   brief notes \cite{mtctv,maruKaon}.
We study  whether the kaonic pasta 
appears in dense neutron star matter, 
and examine the validity  of the Maxwell construction for the
practical use.

One may believe that models taking into account the chiral symmetry are more
realistic for the treatment of the kaon condensation. 
Actually many works have been done  using this approach
\cite{KN86,BLRT20,MTT93,T95,YT00}. 
However, we will exploit  here a non-chiral
model  since the bulk calculations performed within
chiral models do not allow the mixed phase \cite{PREPL00,YT00,kub}. 
Previous investigations of the kaon mixed phase
\cite{RBP00,GS99,PREPL00,CG00,CGS00,NR00} were performed in a
framework of the RMF models. 
Therefore we will also exploit a RMF model for a more straight comparison with the 
previously obtained results.

We also will figure out the surface effect. 
There are few works about this matter so far  \cite{CGS00,NR00,NRprivate}, 
and the finite size effects of the Coulomb interaction and the surface tension are
still not elucidated.  
In Ref.\ \cite{CGS00} authors discussed the surface effect and derived
the surface tension between the normal nuclear matter phase and the kaon
condensed phase by considering two semi-infinite matters 
omitting the Coulomb interaction. 
They suggested that the kinetic terms consisting of space gradients 
of the $\sigma$, $\omega$, $\rho$ mean-fields affect only
slightly the density profiles and the resulting surface tension. 
Therefore they omitted these kinetic terms  while they kept  the kinetic term of 
the kaon field. 
This scale of the kaon field then determines the change of all 
meson fields and the nucleon fields, 
as well as the resulting surface tension, see their Fig.\ 1. 
Then, they find the surface tension which is now
determined by this single length-scale. 
After that, for different geometries minimizing the surface plus the Coulomb energy 
per volume in the structure size one may find the optimum structure size and the geometry.
Note that the Coulomb energy density in their approach 
is given by the bare Coulomb potential decoupling with other fields. 
In reality this approximation does not hold. 
The scale of the change of the kaon field   proves to be of the same order of magnitude as
the scale of the change of the electric field, i.e.\ the Debye screening length. 
A variation of the electric field significantly
affects the kaon distribution and vise versa. 
Therefore, we must discuss the surface effect 
by the kaon field and other meson mean-fields, separately:  
only short-scale fields can decouple
with long-scale fields and thus the contribution of the former fields
may be replaced by the phenomenological surface-tension parameter, 
whereas long-scale fields should be treated explicitly. 

In Ref.\ \cite{NR00} the authors
omitted kinetic terms of the mean $\sigma$, $\omega$, $\rho$ 
fields (neglecting also the corresponding 
surface tension effect) and discussed the charge screening effect. 
By numerical analysis they demonstrated how much the screening effect may affect the
density profiles for the kaon droplet phase.

We consistently incorporate the Coulomb field as the gauge field
coupled with other fields of charged particles and
we keep  kinetic terms of all mean fields. 
We demonstrate that the $\sigma$, $\omega$, $\rho$ 
fields are the short-scale fields.
Thus  we treat consistently the long-range effects, such as screening, 
and the short-range effects and we show their interplay for the kaonic pasta. 
This we believe gives a new insight into the finite size effects in the mixed phase. 
Also by artificial variation of the scale of the short-range
fields we demonstrate effects that can be reduced to  the surface tension.

The paper is organized as follows.
In Sect.\ \ref{Func} we introduce the thermodynamic potential 
and equations of motion for  fields.
In Sect.\ \ref{Results} we numerically solve the Poisson equation and the
mean field equations to find the electric potential profile and the density profiles.
We also derive the EOS and the phase diagram for the kaonic pasta.
In Sect.\ \ref{ChargeK} we discuss the screening effect 
and the surface effect on the kaonic pasta.
Our conclusions are drown in Sect.\ \ref{Sum}. 
Details of the perturbative treatment of the Coulomb effect, 
which we refer to as the ``no Coulomb'' approximation, 
are deferred to Appendix \ref{noCoul}.

\section{Thermodynamic potential and numerical procedure}\label{Func}

\subsection{Thermodynamic potential}

Let us  present our framework to study  kaonic pasta structures.
Following the idea of the density functional theory within the RMF model, we can formulate
equations of motion to  study non-uniform nuclear matter numerically, cf. \cite{refDFT}.
The RMF model with fields of mesons and baryons introduced
in a Lorentz-invariant way is relatively simple for numerical calculations,
but on the other hand it is sufficiently
realistic  to reproduce bulk properties of finite nuclei
as well as the saturation properties of nuclear matter \cite{maru05}.
In our framework, the Coulomb interaction is properly included in
equations of motion for nucleons and electrons and for meson mean fields,
and we solve the Poisson equation for the Coulomb potential $V_{\rm Coul}$
self-consistently with those equations.
Thus the baryon and electron density profiles, as well as the meson
mean fields, are determined in a way fully
consistent with the Coulomb interaction.

We start with the thermodynamic potential for
the system of neutrons, protons, electrons and mesons including kaons
\begin{equation}\label{Omega-tot}
\Omega = \Omega_N+\Omega_M +\Omega_e+\Omega_K.
\end{equation}
The first term
\begin{equation}
\Omega_N  =
  \sum_{a=p,n}
  \int  d^3r
  \left[
  \int_0^{k_{{\rm F},a}}
  { d^3k \over 4\pi^3}
  \sqrt{{m_N^*}^2+k^2}-\rho_a\nu_a
  \right]
\label{OmegaN}
\end{equation}
is the contribution of nucleons with the local Fermi momenta 
$k_{{\rm F},a}({\bf r}); a=n,p$, $m_N^*({\bf r})=m_N-g_{\sigma N}\sigma({\bf r})$
is the effective nucleon mass and $m_N$ is the nucleon mass in the vacuum. 
Nucleons couple with $\sigma$, $\omega$ and $\rho$ mesons and thereby,
\begin{eqnarray}
\nu_n({\bf r})&=&\mu_n-g_{\omega N}\omega_0({\bf r})+g_{\rho N}R_0({\bf r}),\\
\nu_p({\bf r})&=&\mu_p+
{V_{\rm Coul}({\bf r})}-g_{\omega N}\omega_0({\bf r})-g_{\rho N}R_0({\bf r}),
\nonumber
\end{eqnarray}
where $\mu_n$ and $\mu_p$ are neutron and proton chemical potentials
and  $g_{\sigma N}$, $g_{\omega N}$ and  $g_{\rho N}$
are coupling constants between corresponding fields.

The second term  in  (\ref{Omega-tot}) incorporates 
the scalar ($\sigma$) and vector ($\omega_0, R_0$) mean fields,
\begin{eqnarray}
\Omega_M &=& \int   d^3r \Biggl[
  {(\nabla\sigma)^2 + m_\sigma^2\sigma^2 \over2} + U(\sigma) \nonumber\\
 &&{}-{(\nabla\omega_0)^2 + m_\omega^2\omega_0^2 \over2}
  -{(\nabla R_0)^2 + m_\rho^2R_0^2\over2}  \Biggr]  ,
\label{OmegaM}
\end{eqnarray}
where $m_\sigma$, $m_\omega$ and $m_\rho$ are the field masses, and 
$U(\sigma)={1\over3}bm_N(g_{\sigma N}\sigma)^3+{1\over4} c(g_{\sigma N}\sigma)^4$
is the nonlinear potential for the scalar field.

The third term in (\ref{Omega-tot}) contains the contribution of 
the Coulomb field  (described by the potential $V_{\rm Coul}({\bf r})$)
and the contribution of relativistic electrons,
\begin{equation}
\Omega_e = \int d^3r \left[
-{1\over8\pi e^2}(\nabla {V_{\rm Coul}})^2-{(\mu_e-{V_{\rm Coul}})^4\over12\pi^2}
\right],
\end{equation}
where $\mu_e$ is the electron chemical potential.

The last term in (\ref{Omega-tot}) is the thermodynamic potential of the mean $K^-$ meson  field,
\begin{eqnarray}
\Omega_K&=&\!\!\!\int d^3r\left\{
  -{f_K^2\theta^2\over2}\biggl[ 
-{m_K^*}^2+(\mu_K-V_{\rm Coul}\right.
\nonumber\\
  &&\quad \left.
 {} +g_{\omega K}\omega_0+g_{\rho K}R_0)^2 \biggr]+
{f_K^2(\nabla\theta)^2\over2} \right\},
\label{OmegaK}
\end{eqnarray}
where $m_K^*({\bf r})=m_K-g_{\sigma K}\sigma({\bf r})$\, 
is the effective $K^-$ mass, $m_K$ is the kaon mass in vacuum,
$g_{\sigma K}$,  $g_{\omega K}$, $g_{\rho K}$
are the coupling constants, $\mu_K$ is the $K^-$ chemical potential, 
$K=f_K\theta/\sqrt{2}$ is the kaon field and 
$f_K \simeq 93~$MeV is the kaon decay constant
\footnote{%
Note that we consider a linearized $KN$ Lagrangian for simplicity, 
which is not chiral-symmetric.}. 
The kaon charge density $\rho_K$ is expressed in terms of the kaon field $\theta$ as
\begin{equation}
\rho_K = -\left(
\mu_K -V_{\rm Coul} +g_{\omega K}\omega_0 +g_{\rho K}R_0
\right)f_K^2\theta^2.\label{eq:rhoK}
\end{equation}
Temperature $T$ is kept zero in the present study.

For nucleons and electrons we used the local-density approximation, i.e., 
nucleons and electrons are described by their local densities. 
This approximation  has its sense only if the typical length of the change of the 
nucleon density is larger than the inter-nucleon distance.
Derivative terms of the  particle densities can be  incorporated in 
the quasi-classical manner by  the derivative expansion within 
the density functional theory \cite{refDFT}. 
Their contribution to the energy can be reduced to a  surface tension term. 
Here we simply discard those derivative terms, as a first-step approximation. 
Thus we discard the contribution of the nucleon fields to the surface tension assuming
that it is smaller than the corresponding contribution of the meson fields that we retain. 
In the case when we suppress derivative terms of the
nucleon densities they follow changes of the meson $\sigma$, $\omega$, $\rho$ 
mean fields and the kaon and the Coulomb fields that have derivative terms. 
Note that we have fitted our model to properly describe
finite nuclei (see below) without including nucleon derivative terms. 
If we introduced them it would need to re-adjust the model parameters. 
We should  bear in mind that for small structure
sizes quantum effects become prominent. 
For simplicity we disregard these effects. 
Thus within this scheme we may properly describe only rather large-size structures.

\begin{table*}
\caption{
Parameter set used in RMF in our calculation.
}
\begin{center}
\begin{tabular}{ccccccccccccc}
\hline
$g_{\sigma N}$ & 
$g_{\omega N}$ &
$g_{\rho N}$ &
$b$ &
$c$ &
$m_\sigma$ &
$m_\omega$ &
$m_\rho$ &
$f_K (\approx f_\pi)$ &
$ m_K$ &
$g_{\omega K}$ &
$ g_{\rho K}$&
$ U_K(\rho_0)$ \\
\hline\\
6.3935 &
8.7207 &
4.2696 &
0.008659 &
0.002421 &
 400 MeV &
 783 MeV &
 769 MeV &
 93 MeV &
 494 MeV &
$g_{\omega N}/3$&
$g_{\rho N}$&
$-120$ -- $-130$ MeV \\
\hline
\end{tabular}
\end{center}
\end{table*}

\begin{figure*}
  \includegraphics[width=.38\textwidth]{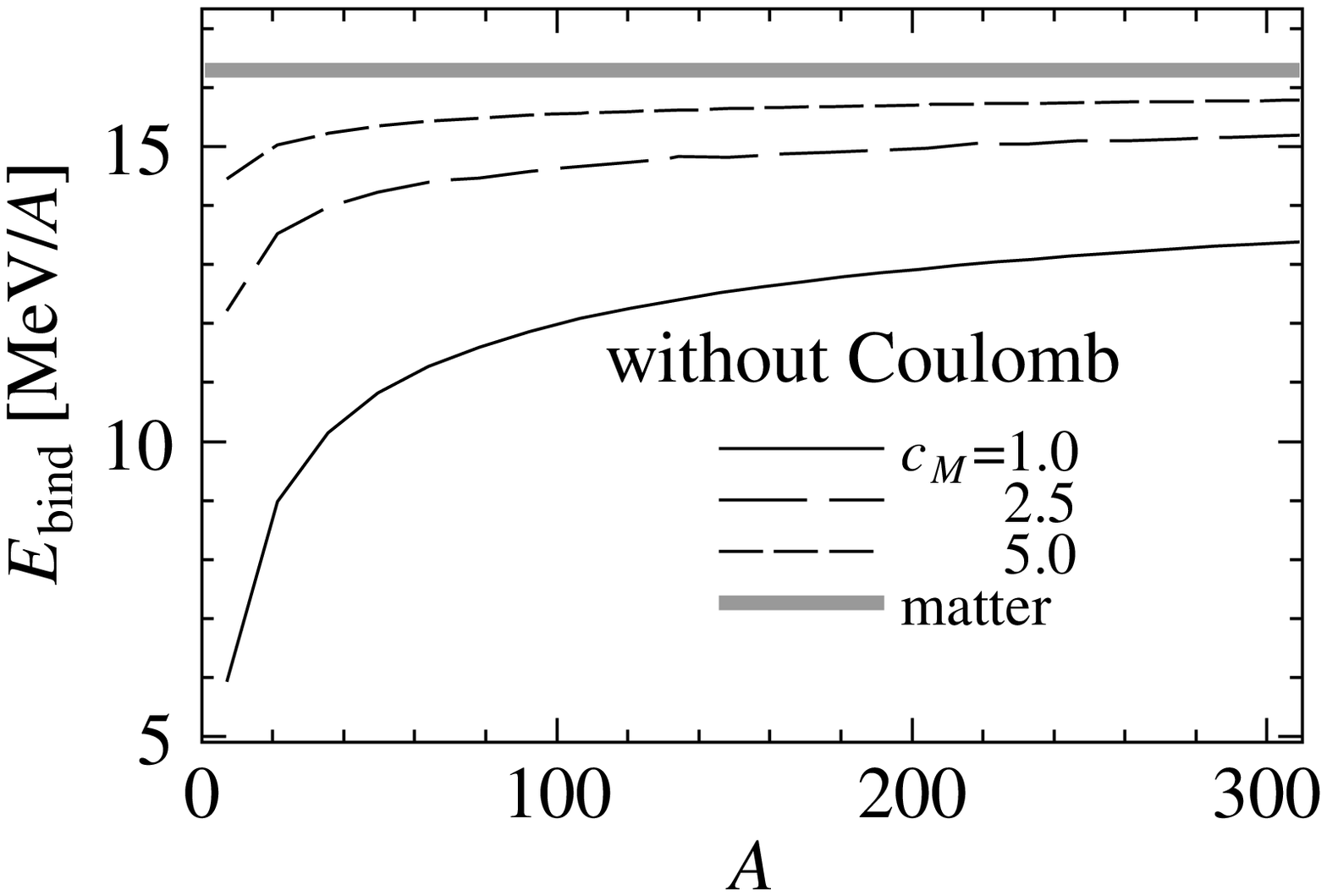}
  \includegraphics[width=.38\textwidth]{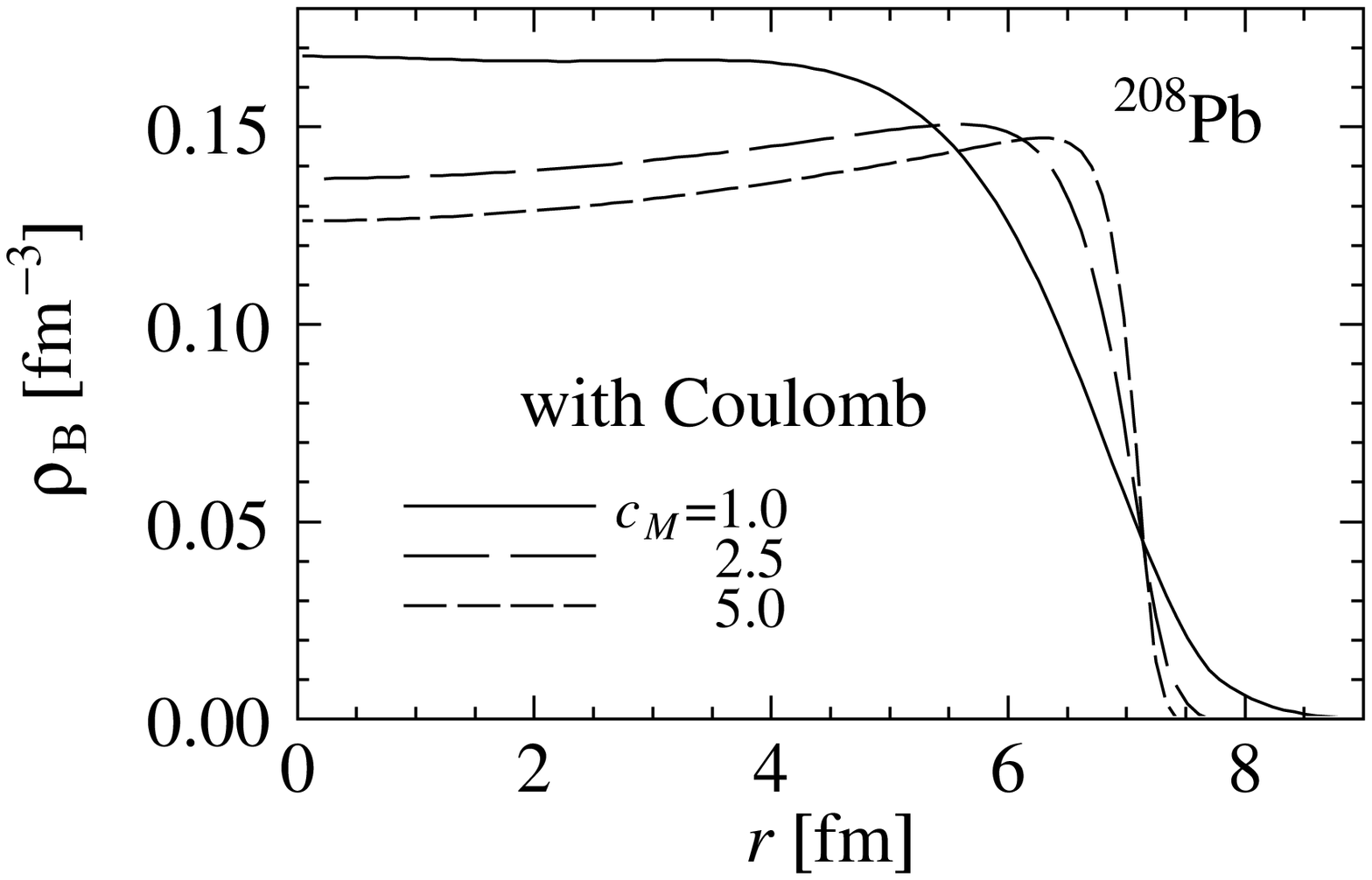}
  \caption{%
  The mass number $A$ dependence of the binding energy of finite nuclei (left)
  and the density profile of $^{208}$Pb nucleus (right).
  Different curves correspond to different meson 
  masses used in the RMF calculation.
  Thick horizontal gray line in the left panel shows the case of the
  nuclear matter for the comparison.
}\label{finiteMs}
\end{figure*}

\subsection{Equations of motion and numerical procedure}

Equations of motion for the mean fields including kaon and 
for the Coulomb potential are obtained
from the variational principle: ${\delta\Omega\over\delta\phi_i({\bf r})}=0$
($\phi_i=\sigma,\omega_0,R_0,\theta ,V_{\rm Coul}$) and
${\delta\Omega\over\delta\rho_a({\bf r})}=0$ ($a=n,p,e$).
These equations read
\begin{eqnarray}
\nabla^2\sigma({\bf r}) &=& m_\sigma^2\sigma({\bf r}) +{dU\over d\sigma}
-g_{\sigma N}(\rho_n^{(s)}({\bf r}) +\rho_p^{(s)}({\bf r})) 
\nonumber\\&&\quad
{}+2g_{\sigma K}m_Kf_K^2 \theta^2({\bf r}),\label{EOMsigma}\\
\nabla^2\omega_0({\bf r}) &=& m_\omega^2\omega_0({\bf r}) -g_{\omega N} (\rho_p({\bf r})+\rho_n({\bf r}))
\nonumber\\&&\quad
{}-f_K^2g_{\omega K}\theta^2({\bf r})(\mu_K-V_{\rm Coul}({\bf r})
\nonumber\\&&\quad
+g_{\omega K}\omega_0({\bf r})+g_{\rho K}R_0({\bf r})),\label{EOMomega}\\
\nabla^2R_0({\bf r}) &=& m_\rho^2R_0({\bf r}) -g_{\rho N}
(\rho_p({\bf r})-\rho_n({\bf r}))
\nonumber\\&&\quad
{}-f_K^2g_{\rho K}\theta^2({\bf r})(\mu_K-V_{\rm Coul}({\bf r})
\nonumber\\&&\quad
{}+g_{\omega K} \omega_0({\bf r})+g_{\rho K}R_0({\bf r})),\label{EOMrho}\\
\nabla^2\theta({\bf r}) &=& \Bigl[{m_K^*({\bf r})}^2-(\mu_K-V_{\rm Coul}({\bf r})
\nonumber\\&&\quad
{}+g_{\omega K}\omega_0({\bf r})+ g_{\rho K}R_0({\bf r}))^2\Bigr]
\theta({\bf r}),\label{EOMtheta}\\
\nabla^2V_{\rm Coul}({\bf r}) &=& 4\pi e^2\rho_{\rm ch}({\bf r}), 
\ \ \ \
\label{EOMpoisson}
\\
\noalign{
the charge density $\rho_{\rm ch}({\bf r})\;=\;\rho_p({\bf r})
+\rho_e({\bf r})+\rho_K({\bf r})$,}\nonumber 
\label{eq:poissonK}\\
\mu_n&=&\mu_B = \sqrt{k_{Fn}({\bf r})^2+{m_N^*({\bf r})}^2}
\nonumber\\&&\quad
{}+g_{\omega N}\omega_0({\bf r})
-g_{\rho N}R_0({\bf r}) \label{EOMmuN},\\
\mu_p&=&\mu_B-\mu_e = \sqrt{k_{Fp}({\bf r})^2+{m_N^*({\bf r})}^2}
\nonumber\\&&\quad
{}+g_{\omega N}\omega_0({\bf r})
+g_{\rho N}R_0({\bf r})-V_{\rm Coul}({\bf r}).
\nonumber\\&&\quad
\label{EOMmuNp}
\end{eqnarray}
The last two equations are the standard relations between the local
nucleon densities and chemical potentials.  
We have assumed that the system is in the chemical equilibrium 
in respect to the weak, electromagnetic and strong interactions
and we introduced the baryon chemical potential $\mu_B =\mu_n$ 
and the charge chemical potential, i.e.\ the electron chemical potential, $\mu_e$, 
according to the corresponding conserved charges. 
Under the same assumption $\mu_K=\mu_e$.

To solve the above coupled equations numerically,
the whole space is divided into equivalent Wigner-Seitz cells.
The geometrical shape of the cell changes as follows:
sphere in three-dimensional (3D) calculation, cylinder in 2D and slab in 1D, respectively.
Each cell is globally charge-neutral and all physical quantities
in the cell are smoothly connected to those of the next cell
with zero gradients at the boundary.
Every point inside the cell is represented by the grid 
points (number of grid points $N_{\rm grid}\approx 100$) and
differential equations for fields are solved by the relaxation method
for a given baryon-number density under constraints of the global charge neutrality. 
Details of the numerical procedure are explained in Ref.\ \cite{maru05}.

\subsection{Parameter set and finite nuclei}

For the study of a non-uniform nuclear matter, the ability to reproduce 
the bulk properties of finite nuclei should be essential.
Parameters of the RMF model are chosen to reproduce saturation properties
of nuclear matter:
the minimum energy per nucleon $-16.3$ MeV at $\rho =\rho_0 \equiv 0.153$ fm$^{-3}$,
the incompressibility $K(\rho_0) =240$ MeV, the effective nucleon  mass
$m_N^{*}(\rho_0)=0.78 m_N$; $m_N =938$ MeV, and the isospin-asymmetry 
coefficient $a_{\rm sym}=32.5$~ MeV.
Coupling constants and meson masses used in our calculation are listed in Table I.

The parameter $g_{\sigma K}$ enters the value of the $K^-$  optical potential $U_K$ 
defined by $U_K = g_{\sigma K}\sigma+g_{\omega K}\omega_0$. 
There have been many works  trying to fix $U_K$ at the saturation density 
from the data on the kaonic atoms \cite{fgb94,fgm99} and from
calculations \cite{wkw96,ro00,ske00,trp02,oor00}, 
but there is still a controversy in its depth. 
We take here a somewhat deep potential, as shown in Table I, 
to compare our results with the previous ones \cite{GS99,CG00,CGS00,NR00}. 
To understand a dependence of the results on the value  $U_K$ 
we further allow  for a variation of it.

We have checked that with the meson masses listed in Table I
($m_{\sigma}=400$ MeV, etc.) and by including the Coulomb
interaction, the binding energies of finite nuclei and the proton fraction,  
as well as the nucleon density profiles are well reproduced, 
except for very light nuclei \cite{maru05}.

We examine in this section how the surface tension could be correctly
incorporated in our RMF calculation.  
We explicitly treat the kaon gradient term in (\ref{EOMtheta})
since, as we show,  the kaon and Coulomb fields  are essentially
coupled and due to that  the scales of changes of these fields
prove to be of the same order of magnitude. 
These are long scales. 
Reference \cite{VC78} using an analytical model 
describing the charge distribution in a pion condensate droplet
has demonstrated that actually there exists only one long scale in the problem.
Namely due to that one cannot reduce the kaon
contribution to a purely surface term in case the Coulomb interaction
is introduced in equations of motion. 

On the other hand the $\sigma$, $\omega$, $\rho$ prove to be short-range fields. 
In order to test  effects of gradient terms of the short-range meson fields in (\ref{OmegaM}) and
(\ref{EOMsigma})--(\ref{EOMrho}) we multiply meson masses
$m_{\sigma}$, $m_{\omega}$ and $m_{\rho}$  by a factor $c_M$. 
We take $c_M =1$ (in realistic case), 2.5 and 5.0. 
For example, with the factor $c_M=2.5$ we get the choice
$m_\sigma$=1000 MeV, $m_\omega$=1958 MeV and $m_\rho$=1923 MeV.
However to obtain appropriate matter properties, we simultaneously fix the ratio 
$g_{\phi N}^2 /m_\phi^2$ ($\phi$=$\sigma$, $\omega$ and $\rho$) for all values of $m_\phi$ 
because  thermodynamic characteristics  depend only on $g_{\phi N}^2 /m_\phi^2$ rather
than on $g_{\phi N}$ and $m_\phi$ separately.

\begin{figure}
  \includegraphics[height=.43\textheight]{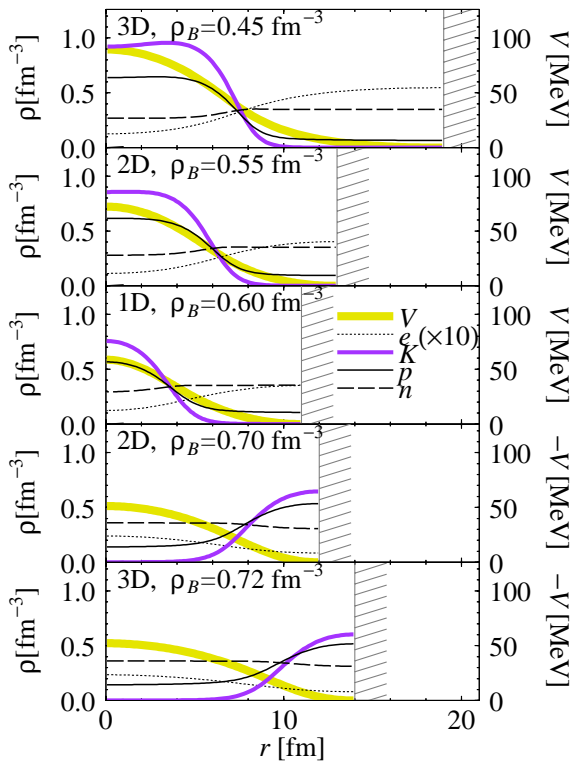}
  \caption{
  (Color online) Density profiles of kaonic structures.
  Here the density does not mean charge-density but number-density 
  of particles.
  $U_k$, the kaon optical potential at the nuclear saturation 
  density, is set to be $-130$ MeV.
  Displayed using the right axis is 
  the Coulomb potential $V_{\rm Coul}$ 
  (written as ``$V$'').
}\label{figProfK}
\end{figure}

First to show the effect of thus simulated surface tension  we
discard the Coulomb interaction (except that we use the global charge
neutrality condition). 
Left panel of Fig.\ \ref{finiteMs} demonstrates
the binding energy of finite nuclei calculated with different meson masses. 
We have checked that for all sets of $\sigma$, $\omega$, $\rho$ 
meson masses (for different $c_M$) the curves approach the value of the
binding energy of the nuclear matter for $A\rightarrow \infty$.
Also by the use of heavy meson masses, the binding energy
of finite nuclei (for finite $A$) 
approaches to that of nuclear matter indicated by a thick gray line. 
This shows that the surface tension is reduced
with the increase of the meson masses, cf. \cite{MS96}. 
Notice that this statement is correct only if we 
fixed the ratio $g_{\phi N}^2 /m_\phi^2$.

Then we allow for the Coulomb interaction. 
Right panel   of Fig.\ \ref{finiteMs} 
shows the baryon density profile of $^{208}$Pb nucleus
obtained with different $c_M$ factors.  
With a larger $c_M$, i.e.\ with a weaker  surface tension, the baryon 
density at the center decreases and that near the surface increases 
due to the Coulomb repulsion of protons.

When we further discuss the kaonic pasta structures in the later section,
we compare results obtained with standard meson masses and those with heavy meson
masses to figure out the effect of the surface tension due to the meson mean-fields.

\section{Results}\label{Results}

The density profile is determined by solving equations of motion 
(\ref{EOMsigma})--(\ref{EOMmuNp}).
For each baryon density, the cell radius and the geometrical dimension 
are chosen to minimize the energy. 
If Glendenning's claim were correct, the structured mixed phase would 
develop in a broad density range from well below to well above 
the critical densities determined by the Maxwell construction. 
In this density interval the matter should exhibit the structure change, 
similar to the nuclear ``pasta'' phases \cite{maru05}: 
the kaonic droplet, rod, slab, tube, bubble.
We observe such structures in our calculation as well. 

Figure \ref{figProfK} displays typical density profiles and the Coulomb potential. 
The horizontal axis is the distance from the cell
center and the hatch shows the cell boundary. 
The symbols ``3D'' (three dimensional) etc.\ indicate the 
dimensionality of the geometry.
 From the top of the figure, the matter structures correspond 
to kaonic droplet, rod, slab, tube, and bubble.
The neutron distribution proves to be rather flat. 
The Coulomb potential is given by imposing the gauge condition. 
Here we fix the gauge by the condition,  $V_{\rm Coul}(R_{\rm cell})=0$
for all the cases. 
Note that the maximum value of the Coulomb potential becomes rather large.

\begin{figure}
  \includegraphics[height=.36\textheight]{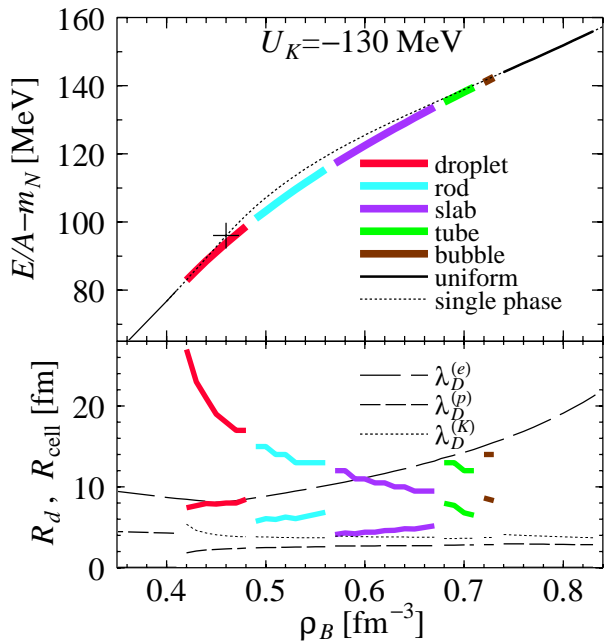}
  \caption{(Color online)
  Top: Binding energy per nucleon of the
  nuclear matter in the beta equilibrium.
  The dotted line below the cross shows the uniform normal nuclear matter 
  and above the cross, the uniform kaonic matter.
  Bottom: Structure size $R_d$ (thick curves below)
  and cell size $R_{\rm cell}$ (thick curves above).
  Compared are the Debye screening lengths of electron, proton and kaon.
  The Debye screening lengths are calculated using the 
  explicit dependence of $\mu_a$ on $\rho_a$.
}\label{figKEOS}
\end{figure}

In the upper panel of Fig.\ \ref{figKEOS} we depict the energy per
nucleon of the matter. 
The dotted line indicates the case of single phase 
(if one assumes absence of the mixed phase).
In this case the uniform matter consists of
normal nuclear matter for  undercritical densities
and the kaonic matter for overcritical densities.
The cross  on the dotted line ($\rho_B \simeq 0.46$~fm$^{-3}$)
shows the critical density, i.e.\ the point where kaons begin 
to condensate in the case of single phase.
Pieces of solid curves, on the other hand, 
indicate  energetically favored structures. 
Droplets begin to appear for $\rho_B >0.41$~fm$^{-3}$ 
smoothly decreasing the energy of the system. 
The mixed phase disappears for $\rho_B >0.74$~fm$^{-3}$.
The occurrence of the kaonic pasta structures, 
i.e.\ kaonic droplet, rod, slab, tube, and bubble, 
results in a  softening of the matter (the energy decreases with their appearance).

In the lower panel of Fig.\ \ref{figKEOS} we plot the structure sizes 
(for droplet, rod, slab, tube and bubble) $R_d$ and the corresponding cell sizes $R_{\rm cell}$. 
The structure size $R_d$ is calculated by the use of 
the averaged density and its spacial fluctuation as in Ref.\ \cite{maru05}.
We find that at the onset density the radius of the cell is infinitely large
in case of the full calculation. 
The corresponding steep increase of $R_{\rm cell}$ 
with decreasing density is clearly seen in the figure.

To roughly estimate the Debye screening length $\lambda_D^{(a)}$
(see Appendix \ref{debye} for detail) 
let us for simplicity use  only  explicit dependence of 
$\left<\rho_a\right>$ on $\mu_a$ in (\ref{Deb}), i.e., let us replace  
$\displaystyle {d\left<\rho_a\right>\over d\mu_a}$
by $\displaystyle {\partial\left<\rho_a\right>\over \partial\mu_a}$, 
hence $\displaystyle {\partial\left<\rho_e\right>\over \partial
\mu_e}\approx-{\mu_e^2\over \pi^2}$, 
$\displaystyle {\partial\left<\rho_p\right>\over \partial\mu_p}\approx
{\left<k_{Fp}\right> \sqrt{{\left<k_{Fp}\right>}^2+{m_N}^2}\over \pi^2}$ 
for electron and proton, and 
$\displaystyle {\partial\left<\rho_K\right>\over \partial
\mu_K}\approx \left<-f^2\theta^2\right>$ for kaon.
Here, the bracket ``$\langle ... \rangle$'' indicates an averaging
over an appropriate region (cf.\ Appendix \ref{debye}). 
The dashed lines and the dotted line in the lower panel of Fig.~\ref{figKEOS}
show thus treated partial contributions to the Debye screening lengths of 
the electron, proton and kaon,
$\lambda_{D}^{(e)}$, $\lambda_{D}^{(p)}$, 
and $\lambda_{D}^{(K)}$, respectively. 
We see that in most cases $\lambda_{D}^{(e)}$ is less than the cell size
$R_{\rm cell}$ but it is larger than the structure size $R_d$.
The proton Debye length $\lambda_{D}^{(p)}$ 
and the kaon Debye length $\lambda_{D}^{(K)}$, on the other hand, 
are always shorter than $R_{\rm cell}$ and the structure size $R_d$.
Now we are able to estimate the value
of the  resulting total Debye screening length $\lambda^{(\rm tot)}_D $. 
For example  in the case of $\rho_B=0.55$ fm$^{-3}$  we obtain
$\lambda^{(\rm tot)}_D\simeq 3$~fm. 

On the other hand the typical length of the Coulomb potential is $\sim 5$ fm 
for $\rho_B=0.55$ fm$^{-3}$ (see Fig.\ \ref{figProfK}).
This value is longer than the above estimated Debye screening length
due to the implicit dependence of $\rho_a$ on $\mu_a$.

 From this comparison we conclude that a consistent inclusion of 
 the Coulomb screening effect is not a  trivial problem and it has 
really important consequences for the correct description
of density profiles of protons and kaons 
(see Fig.~\ref{figKprofcompare} below). 

So far we have presented results for the $K^-$ optical potential 
in nucleus equal to $U_K=-130$ MeV.
With a lower value $U_K=-120$ MeV,
the density range of the kaonic pasta is narrowed:
from $\rho_B=0.49$ to 0.71 fm$^{-3}$ instead of from 
$\rho_B=0.41$ to 0.74 fm$^{-3}$ in the case of $U_K=-130$ MeV.
The effect  of the kaon pasta structures on the EOS 
(the energy difference between single phase calculation and
full calculation) becomes still smaller.
The  kaon and proton densities in a droplet become 20 -- 30 \%
lower than those for the case $U_K=-130$ MeV,
and the neutron density increases.
Other  features remain qualitatively the same.

\section{Charge screening and  surface effects}\label{ChargeK}

\subsection{Charge screening effect}

To demonstrate the charge screening effects we compare our results with
those given by a ``perturbative'' treatment of the Coulomb interaction often used in the
literature, ``no Coulomb'' calculation (see Appendix \ref{noCoul} for details). 
The electric potential is discarded in equations of motion 
(\ref{eq:rhoK})--(\ref{EOMtheta}), (\ref{EOMmuN}), (\ref{EOMmuNp}) 
which determine the density profiles.  
The Coulomb energy is then added to the total energy by using the
charge density profile thus determined to find the optimal value
with respect to the cell size $R_{\rm cell}$.
\begin{figure}
  \includegraphics[height=.36\textheight]{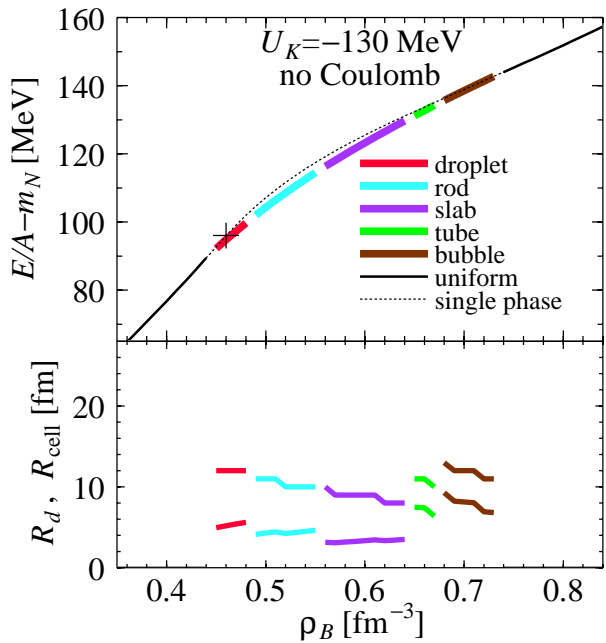}
\caption{(Color online)
  Top: the  binding energy per nucleon of the
  nuclear matter in  beta equilibrium.
  The electric potential is discarded determining the density profile 
  and added evaluating the energy.
  Bottom: the structure size $R_d$ (thick curves below)
  and the cell size $R_{\rm cell}$ (thick curves above).
}\label{figKEOSNC}
\end{figure}
Comparing Figs.\ \ref{figKEOS} and ~\ref{figKEOSNC}, 
we see that the density range of the mixed phase is
more narrow in the case of the ``no Coulomb'' calculation than 
in the full calculation, while the energy gain is almost the same. 
A remarkable difference is seen in the cell radii, 
especially near the onset density of kaonic pastas, for $\rho_B<0.5~$fm$^{-3}$.
The cell size given by the full calculation is always larger than that
given by the ``no Coulomb'' calculation.  

\begin{figure}
  \includegraphics[height=.36\textheight]{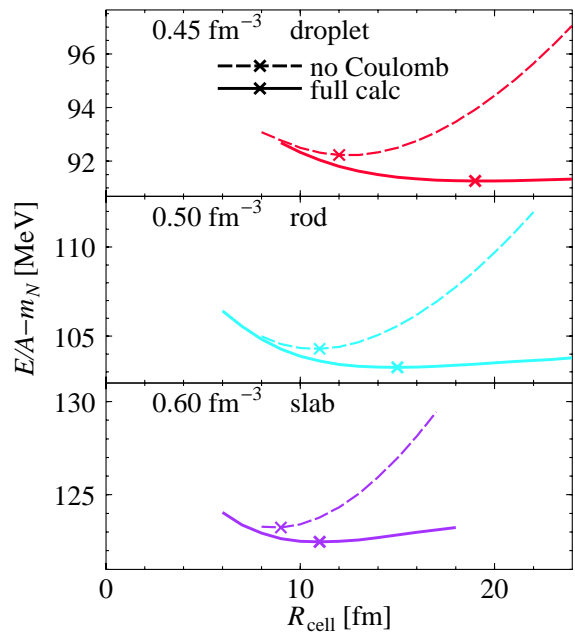}
\caption{(Color online)
  The cell size $R_{\rm cell}$ dependence of the energy per nucleon.
  Crosses on the curves indicate the minimum points.
}\label{figRE}
\end{figure}

To elucidate the screening effect, we depict the $R_{\rm cell}$ 
dependence of the energy per nucleon in Fig.~\ref{figRE} in both cases.
In the full calculation a large cell with a small volume fraction $f\equiv
(R_d/R_{\rm cell})^3$ appears near the onset density,
which situation is close to the uniform single phase.
On the other hand, a smaller cell appears near the onset density
in the ``no Coulomb'' calculation.
The energy gain is higher in the full calculation. 
The cell radii obtained in both calculations deviate 
essentially for droplets, less for rods, still less for slabs, etc. 

\begin{figure}
  \includegraphics[height=.38\textheight]{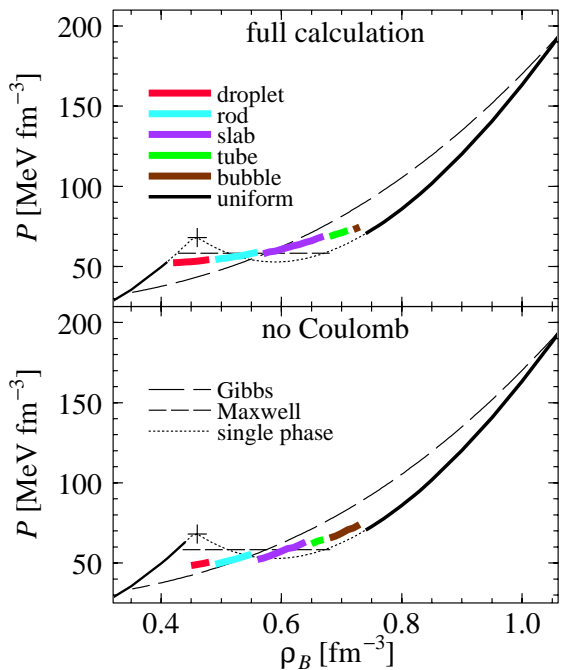}
\caption{(Color online)
  The pressure versus the baryon number density.
}\label{figRhoP}
\end{figure}

The differences of both calculations are also seen 
from Fig.~\ref{figRhoP}, where we present the pressure 
as a function of the baryon-number density.
In the case of the full calculation the pressure changes continuously.
In the case of the ``no Coulomb'' calculation there arises
a jump in the pressure at the onset density at which the droplets begin to appear.
The pressure jump clearly shows that a large structure
change occurs at the onset density. 
The discontinuity arises as an artifact of the ``no Coulomb'' calculation. 
The Coulomb energy is overestimated in this case for large values of 
$R_{\rm cell}$, compared to that for the full calculation, where occurs the screening effect. 
Indeed, the Coulomb energy per droplet grows  proportionally to $R_{\rm cell}^2$
in the case of the ``no Coulomb'' calculation since the bare
Coulomb interaction has an infinite range. 
On the other hand in the full calculation there appears another length scale, 
the Debye screening length, and thereby the screened Coulomb interaction is
no longer scale-invariant. 
Thus when the structure radius is significantly larger than the
minimal screening length the Coulomb energy contribution  becomes ineffective. 

In  Fig.~\ref{figRhoP}, we also depict the pressure when  
the Gibbs conditions are applied for two semi-infinite
matters disregarding the Coulomb interaction (indicated by ``Gibbs'') 
and that given by the Maxwell construction (indicated by ``Maxwell'').
We see that the pieces of solid curves  lie between ``Gibbs'' and ``Maxwell''. 
The full calculation case is more similar to the one given 
by the Maxwell construction.

\begin{figure}
\includegraphics[width=.49\textwidth]{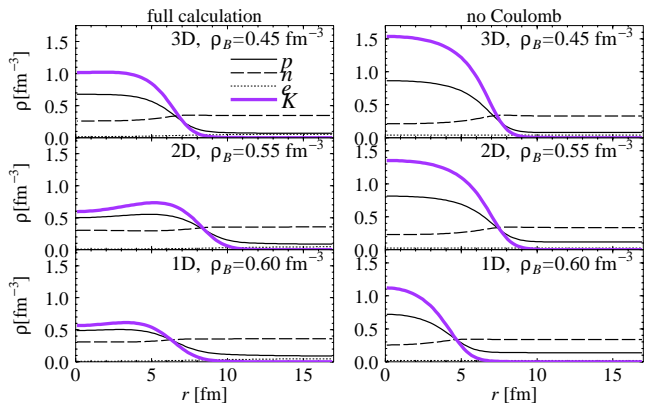}
  \caption{(Color online)
  Comparison of density profiles of kaonic matter in full
  calculation  and in ``no Coulomb'' case.
  The cell size, $R_{\rm cell}=17$~fm, is not optimized 
  since the optimum values would be different for different 
  treatments of the Coulomb interaction.
}
\label{figKprofcompare}
\end{figure}

To further demonstrate the charge screening effect on the kaonic pasta, 
We compare the density profiles 
obtained by the full and ``no Coulomb'' calculations
in Fig.~\ref{figKprofcompare}. 
In case of the full calculation the difference between 
the negative charge density (of kaons and electrons) and 
the positive charge density of protons  is smaller,
demonstrating that the profiles are more close to those given by the
local charge neutrality.
These results also suggest that the Maxwell construction is effectively 
justified in the full calculation case owing to the charge screening effects.  
Also we see that the absolute value of the kaon charge density is 
substantially larger in the ``no Coulomb'' case.
This gives us an additional argument for the essential
coupling of the kaon and Poisson equations. 
Modification of the Coulomb term in the kaon equation significantly affects the 
charged kaon distribution whereas changes in the proton, neutron and
electron distributions  are more smooth.

\subsection{Surface effects}

Next we investigate the effect of the surface tension caused by the
meson mean-fields on the kaonic pasta structures. 
The previous calculation by Norsen and Reddy \cite{NR00,NRprivate} 
was done discarding the gradient terms of $\sigma$, $\omega$ and $\rho$ mesons.
The calculation with infinitely heavy meson masses would correspond to it.

\begin{figure}[h]
  \includegraphics[height=.36\textheight]{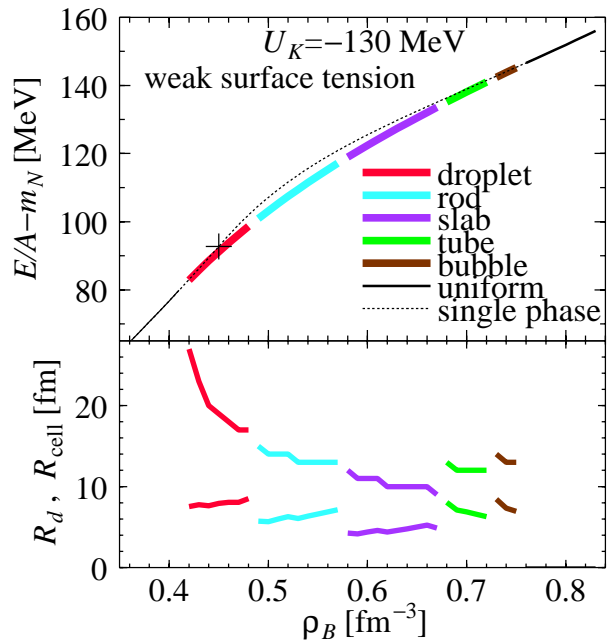}
\caption{(Color online)
  Top:  Binding energy per nucleon of
  the nuclear matter in  beta equilibrium.
  Bottom:  Structure size $R_d$ (thick curves below)
  and the cell size $R_{\rm cell}$ (thick curves above).
}\label{figKEOSms}
\end{figure}

Figure \ref{figKEOSms} shows the same quantities as  Fig.\ \ref{figKEOS},
EOS and the structure size, but now for the case of an artificially suppressed 
surface tension (for $c_M=5.0$).
Comparing Figs.\ \ref{figKEOSms} and \ref{figKEOS}, we see that
there is almost no difference in the EOS. 
However, the density range of pasta structure is slightly broader for 
the case of the weaker surface tension. 
The reason for a similarity of Figs.\ \ref{figKEOSms} and \ref{figKEOS} is as follows:
Even without the contribution of gradient terms of
$\sigma$, $\omega$ and $\rho$ mesons (that corresponds to
infinitely massive mesons, however for fixed $g_{\phi N}^2 /m_\phi^2$), 
there is still a contribution from the kaon gradient term in Eq.\ (\ref{OmegaK}). 
In our case the kaon and the Coulomb fields have similar typical length scales. 
The Coulomb term is mainly balanced by the kaon term, 
and the structure size is actually determined by
the minimum  of the Coulomb plus kaon kinetic energy.
The pasta structure is realized by the balance between 
the surface tension and the Coulomb repulsion. 
In this sense, $\sigma$, $\omega$ and $\rho$ mesons have less contribution
to the surface tension compared to kaon in the case of kaonic pasta structure,
although the clear distinction of the Coulomb repulsion 
and the surface tension is difficult due to
the screening effects and the large surface diffuseness.

\section{Summary and concluding remarks}\label{Sum}

We have discussed the effects of the charge screening and the
surface tension on the kaonic pasta.
Since the kaon mixed-phase appears at high-densities, we see that changes are
more remarkable than for the ``nuclear pasta'' at subnuclear densities \cite{maru1}.
We found that the density range of the structured mixed
phase is largely limited by the charge screening and thereby the
phase diagram becomes similar to that given by the Maxwell construction.
Although the importance of such a treatment has been demonstrated
for the hadron-quark matter transition \cite{voskre,emaru1}, one
of our new findings here is that we can figure out the role of the
charge screening effect without introducing any ``artificial''
input of the surface-tension parameter. 
For this aim in our study we have used the  relativistic  mean field approach.

On the other hand we remind that the bulk calculation, 
where finite size effects (the Coulomb interaction and the surface tension) 
are not included, does not produce the mixed phase 
in the chiral models \cite{PREPL00,YT00,kub}. 
Thus it should be interesting to perform
the same self-consistent study with properly included finite size effects, 
as we have done above, but within a chiral model. 

In application to the newly formed neutron stars in supernova explosions, 
finite temperature and neutrino trapping effects become important, 
as well as the dynamics of the first order phase transition with 
formation of the structures.
It would be interesting to extend our framework to include these effects.

We focused our study on the $s$-wave kaon condensation. 
However it would be interesting to extend it 
to the possibility of the $p$-wave kaon condensation.

It might be also interesting to apply our framework to the problem of 
kaonic nuclei \cite{ay02,d04,k99,i04}, similar to the description of finite nuclei. 
Within our framework we can construct a consistent description of kaonic
nuclei by taking into account the Coulomb interaction as well as
the $KN$ interaction \cite{maru052}.

\section*{Acknowledgments}

We thank S.~Reddy for discussions and communications.
This work is partially supported by the Grant-in-Aid for the 21st Century COE
``Center for the Diversity and Universality in Physics'' from
the Ministry of Education, Culture, Sports, Science and Technology of Japan. 
It is also partially supported by the Japanese
Grant-in-Aid for Scientific Research Fund 
of the Ministry of Education, Culture, Sports, Science and Technology (13640282, 16540246). 
The work of D.N.V.\ was also supported in part by the Deutsche
Forschungsgemeinschaft (DFG project 436 RUS 113/558/0-2).

\appendix

\section{``No Coulomb'' calculation}\label{noCoul}

We briefly describe here   the ``no Coulomb'' approximation, which has been frequently 
used in the bulk calculations. 
To begin with we present the thermodynamic potential (\ref{Omega-tot}) as
\begin{equation}
\Omega(V_{\rm Coul})=\Omega_{\rm matter}(V_{\rm Coul})-\int d^3r\frac{1}{8\pi e^2}(\nabla V_{\rm Coul})^2,
\label{a}
\end{equation}
where we isolated the electric field contribution and $\Omega_{\rm matter}(V_{\rm Coul})$ 
summarizes other baryon, electron, kaon and mean-field contributions.
Expanding $\Omega(V_{\rm Coul})$ around a reference value $V_{\rm Coul}=V_{\rm 0}=const.$ we have
\begin{eqnarray}
\Omega(V_{\rm Coul})&=&\Omega_{\rm matter}(V_0) \nonumber\\
&&\hspace*{-15mm}{}+\int d^3r \left.\frac{\delta\Omega_{\rm matter}(V_{\rm Coul})}{\delta V_{\rm Coul}(r)}
\right|_{V_0}\left(V_{\rm Coul}(r)-V_0\right) \nonumber\\
&&\hspace*{-15mm}{}-\int d^3r\frac{\left(\nabla V_{\rm Coul} \right)^2}{8\pi
  e^2}+O\left[ (V_{\rm Coul}(r)-V_0)^2\right].
\label{b}
\end{eqnarray}
The non-zero value of $V_0$ shifts the value of the charge
chemical potential in such a way  that only 
the combination $\mu-V_{\rm Coul}+V_0$ is meaningful due to the gauge invariance. 
To be specific  in the paper body we performed
calculations in the ``no Coulomb'' case  with $V_0=0$.
Since $\Omega_{\rm matter}(V_{\rm Coul})$ 
includes $V_{\rm Coul}$ in a gauge invariant way, it can be further rewritten as
\begin{eqnarray}
\Omega(V_{\rm Coul})&=&\Omega_{\rm matter}(V_0)
-\int d^3r\rho^{\rm ch}(V_0)\left(V_{\rm Coul}(r)-V_0\right)
\nonumber\\
&&\hspace*{-10mm}{}-\int d^3r\frac{\left(\nabla V_{\rm Coul}\right)^2}{8\pi e^2}+O\left((V_{\rm Coul}(r)-V_0)^2\right).
\label{c}
\end{eqnarray}
Using the total charge neutrality condition, it eventually reads
\begin{equation}
\Omega(V_{\rm Coul})\equiv\Omega^{\rm NC}(V_{\rm Coul})+O\left((V_{\rm Coul}(r)-V_0)^2\right),
\end{equation}
with
\begin{eqnarray}
\Omega^{\rm NC}(V_{\rm Coul})&=&\Omega_{\rm matter}(V_0)
-\int d^3r\rho^{\rm ch}(V_0)V_{\rm Coul}(r)\nonumber\\
&&{}-\int d^3r\frac{\left(\nabla V_{\rm Coul}\right)^2}{8\pi e^2}.
\label{d}
\end{eqnarray}
The Poisson equation for $V_{\rm Coul}$ is given by
\begin{equation}
\nabla^2 V_{\rm Coul}=4\pi e^2\rho^{\rm ch}(V_0),
\label{e}
\end{equation}
while the equations of motion for other fields and density profiles, 
$f_\alpha({\bf r})\equiv\phi_i({\bf r}), \rho_a({\bf r})$, read 
\begin{equation}
\frac{\delta\Omega_{\rm matter}(V_0)}{\delta f_\alpha({\bf r})}
-\frac{\partial\rho^{\rm ch}(V_0)}
{\partial f_\alpha({\bf r})}V_{\rm Coul}(r)=0.
\label{f}
\end{equation}
In (\ref{f}) one can neglect the second term of the electromagnetic origin 
compared to the first term related to the strong interaction. 
Then one has
\begin{equation}
\frac{\delta \Omega_{\rm matter}(V_0)}{\delta f_\alpha({\bf r})}\simeq 0,
\label{g}
\end{equation}
Using Eq.~(\ref{e}), we can recast Eq.~(\ref{d}) into the form,
\begin{eqnarray}
\Omega^{\rm NC}(V_{\rm Coul})&=&\Omega_{\rm matter}(V_0)
-\frac{1}{2}\int d^3r\rho^{\rm ch}(V_0)V_{\rm Coul}(r)\nonumber\\
&=&\Omega_{\rm matter}(V_0)
+\int d^3r\frac{\left(\nabla V_{\rm Coul}\right)^2}{8\pi e^2}.
\label{h}
\end{eqnarray}
Equations (\ref{e}), (\ref{g}) and (\ref{h}) are used in the ``no Coulomb'' calculation. 
The second term in Eq.~(\ref{h}) (let's call it  the Coulomb energy) gives a
positive (repulsive) contribution.

The corresponding total energy $E^{\rm NC}$ is then given 
by the Legendre transformation from $\Omega^{\rm NC}$,
\begin{equation}
E^{\rm NC}=\Omega^{\rm NC}+\mu_B\sum_{a=n,p}\int d^3r\rho_a
\end{equation}
with
\begin{equation}
\rho_a=\frac{\partial\Omega_{\rm matter}(V_0)}{\partial\mu_a}.
\end{equation}
In order to proceed  beyond the ``no Coulomb'' approximation one should 
perform the expansion in (\ref{b}) at least including the quadratic terms. 
With these terms one arrives at the
linearized Poisson equation for the Coulomb potential that incorporates 
the Debye screening, cf. \cite{voskre,voskre1}. 

\section{Debye screening length}\label{debye}
To roughly estimate the screening effect we can introduce 
the Debye screening lengths as follows.
A linearized Poisson equation for the Coulomb potential
$\delta V_{\rm Coul} = V_{\rm Coul} -\mu_e -V_0$, 
where $V_0$ is the reference constant value,
takes the form 
\footnote{%
Note that only the combination $V_{\rm Coul}-\mu_e -V_0$ 
matters due to the gauge invariance.}
\begin{eqnarray}\label{lin}
&&\Delta \delta V_{\rm Coul} +\left(\lambda^{(\rm tot)}_D\right)^{-2}\delta V_{\rm Coul}=0,\\
&&\left(\lambda^{(\rm tot)}_D\right)^{-2}= \sum_{a=e,p,K} \left(\lambda^{(a)}_D\right)^{-2}, \\
&&\lambda^{(a)}_D= \left(4\pi e^2{d\left|\left<\rho_a\right>\right| \over d\mu_a}\right)^{-1/2}.
\label{Deb}
\end{eqnarray}
The partial contributions $\lambda^{(a)}_D$, the Debye screening lengths,
are dependent on the region. 
To find  typical values characterized screening inside the structure 
we average $\rho_a$ over the region where the relevant charge density is non-zero. 
Thus the averaging region for proton and kaon 
is inside the lump 
and that for electron is the cell.

To obtain  correct values of the Debye screening lengths, the implicit dependence
of $\left<\rho_a\right>$ on $\mu_a$ through the change of mean fields and the Coulomb potential
should be taken into account. 
We will see that it is especially in the case of the kaon Debye term. 
Indeed, the linearization procedure should be performed in all fields and
the linearized Poisson equation is as follows:
\begin{eqnarray}\label{lina}
&&\Delta \delta V_{\rm Coul} +a_{VV}\delta V_{\rm Coul}+a_{V\sigma} 
\delta\sigma\\
&&+a_{V\omega} \delta\omega_0 +a_{V\rho} \delta R_0 +a_{V\theta} 
\delta\theta \nonumber\\
&&+a_{Vn} \delta \rho_n +a_{Vp} \delta \rho_p 
=0,\nonumber
\end{eqnarray}
where $a_{Vi}$ are density dependent coefficients.
The fields $\delta\sigma$, $\delta\omega_0$, $\delta R_0$, $\delta\theta$ 
are determined by their own equations of motion having  similar forms.
Also variation of equations for $\mu_n$ and $\mu_p$ yields equations for
$\delta\rho_n$ and $\delta\rho_p$, $\mu_e$ is unambiguously expressed via
the neutron and proton chemical potentials, therefore we did not present the
corresponding term in (\ref{lina}). 
Neglecting derivative terms for all the fields
except for the Coulomb field in the system of equations similar to 
(\ref{lina}) one can express $\delta\sigma$, $\delta\omega_0$, $\delta R_0$,
$\delta\rho_n$ and $\delta\rho_p$ as functions of the only one $\delta V_{\rm Coul}$ variable 
and one finally arrives at Eq.\ (\ref{lin}). 
However such a procedure is legitimate only for short-scale fields, 
$\delta\sigma$, $\delta\omega_0$, $\delta R_0$,
$\delta\rho_n$ and $\delta\rho_p$ but not for the kaon field $\delta\theta$
since the later field proves to be a long-scale field.
The value $\Delta \delta\theta$ entering the kaon equation of motion 
cannot be omitted thereby. 


\end{document}